\documentclass[letter,twocolumn]{jpsj2}
\usepackage{amsmath}

\newcommand{\sgn}{\mathrm{sgn}}
\newcommand{\jc}{j_{\mathrm{K}}^{x}}

\title{Nonequilibrium Microscopic Distribution of Thermal Current in Particle Systems}

\author{Satoshi \textsc{Yukawa},\thanks{E-mail: yukawa@ess.sci.osaka-u.ac.jp}
Takashi \textsc{Shimada},$^{1}$ Fumiko \textsc{Ogushi},$^{1,2}$ and Nobuyasu \textsc{Ito}$^{1}$
}
\inst{
Department of Earth and Space Science, Graduate School of Science, Osaka University, Toyonaka 560-0043, Japan,\\
$^{1}$ Department of Applied Physics, School of Engineering, The University of Tokyo, Bunkyo 113-8656, Japan\\
$^{2}$ Theoretical Biochemistry Laboratory, RIKEN Advanced Science Institute,  Wako 351-0198, Japan}
\abst{%
A nonequilibrium distribution function of microscopic thermal current 
is studied by a direct numerical simulation in a thermal conducting steady state of particle systems. 
Two characteristic temperatures of the thermal current are investigated on the basis of the distribution. 
It is confirmed that the temperature depends on the current direction; 
Parallel temperature to the heat-flux is higher than antiparallel one. 
The difference between the parallel temperature and the antiparallel one is proportional to a macroscopic temperature gradient.
}
\kword{nonequilibrium steady state, thermal transport, current distribution, particle system}

\begin{document}

\maketitle 

The characterization of nonequilibrium steady state (NESS) is one of the important issues of nonequilibrium statistical mechanics. 
For a linear nonequilibrium regime, several methods describe the nature of NESS.
The linear response theory gives a method of calculating response coefficients from equilibrium fluctuations. 
An approach using the Boltzmann equation enables us to obtain response coefficients through a nonequilibrium distribution. 
However, in general situations such as that in a nonlinear nonequilibrium regime, we have no clear nor general methods for such a description. 

In the following, NESS is analyzed on using a distribution function, because a distribution 
function implies direct information on NESS.
In particular, we investigate a microscopic distribution of the energy current of particle systems in a steady thermal-transport state. 
A thermal-transport system is a simple and well-studied example of nonequilibrium steady states.\cite{LLP03} 
Microscopic energy current is carried by a single particle in a particle system. 
If no net flow of particles exists in the system, the energy current corresponds to microscopic thermal current. 
Gathering microscopic thermal current in an appropriate cross section, we obtain macroscopic thermal current. 
From the viewpoint of nonequilibrium statistical mechanics, macroscopic thermal current has less information than 
microscopic thermal current. Generally, its distribution is Gaussian because of the central limit theorem and has only 
two parameters: average and variance.

Let us consider an equilibrium case of the microscopic distribution before a nonequilibrium case. 
For a particle system in three-dimensional space, an $x$-component of the kinetic part of the thermal current $\jc$ is defined as\cite{McL88}
\begin{equation}
\jc \equiv  \dfrac{p_{x}^{2}+p_{y}^{2}+p_{z}^{2}}{2m} \dfrac{p_{x}}{m} \enspace,
\end{equation}
with a mass $m$ and a momentum $ \mathbf{p} = \left( p_{x},p_{y},p_{z}\right).$
The distribution of $\jc$ is obtained by taking the thermal average of $\delta (j-\jc)$ with a Maxwellian distribution of temperature $T$. 
(Hereafter, we take the Boltzmann constant to be unity.)  
It is given by
\begin{align}
P_{\mathrm{eq}}(j) & \equiv \left\langle \delta \left( j - \jc\right)\right\rangle_{\mathrm{eq}}  = \sqrt{\dfrac{\mu}{2\pi }} \mathrm{E}_{1}
\left( z \right)
\label{eq:eqdist}
\end{align}
with a normalization $\int_{-\infty}^{\infty}dj \, P_{\mathrm{eq}}(j)=1$,
where $\mu$ is a temperature-scaled mass parameter defined by $\mu\equiv m \slash T^{3}$, which has the dimension of the inverse 
square of thermal current, and a dimensionless thermal current with a fractional power $z\equiv  \left(\mu \slash 2 \right)^{{1/3}}\lvert j  \rvert^{2/3}$.
$\mathrm{E}_{1}(z)$ is the $\mathrm{E}_{n}$-function with 
$n=1$ defined by\cite{AS72} 
\begin{equation}
\mathrm{E}_{1}(z) = \displaystyle \int_{1}^{\infty} dx \dfrac{e^{-zx}}{x} 
\enspace ,
\end{equation}
which is related to the exponential integral function $\mathrm{E}_{1}(z) = -\mathrm{Ei}(-z)$.
In this letter, we treat only the kinetic part, although a distribution of the potential part of the thermal current can also be evaluated\cite{SOI07}. 
The equilibrium distribution eq.~(\ref{eq:eqdist}) has a log divergence, $-\ln z$, near $j \simeq 0$
and an exponential tail with an algebraic correction, $ \sim e^{-z}\slash z  \enspace (z \to \infty).$

Let us consider a nonequilibrium distribution of $\jc$.
For a small thermal current $j \simeq 0$, it should behave like an equilibrium distribution with a 
local equilibrium temperature, that is, it has a log divergence, because the small current is contributed 
by low-energy particles that are well thermalized locally.
High-energy particles contribute to the negative and positive tails of the distribution.
They come from far places from the observation point without scattering by other particles. 
Therefore, these high-energy particles have information on these places. 
When these places are separated by a scale larger than the mean free path, 
the tails of the distribution of the thermal current may have some information on local equilibria. 
If these local equilibria are characterized by two typical equilibria that are separated by the scale of the mean free path, 
it means that the apparent temperatures of the tails are different from the local temperature of the observation point, 
and that the two temperatures are related to a thermal gradient and a scale of the mean free path.
If the temperature gradient $\partial T \slash \partial x$ is negative, a large positive $x$ 
component of thermal current has  information on the  high-temperature local equilibrium and a large negative $x$ 
component has information on the low-temperature local equilibrium.

The above picture predicts that the tails of the distribution of the current can be described by two 
other local equilibrium distributions with different temperatures. For an ideal gas system, this is trivial. 
In this letter, we study this picture for dilute Hertzian-sphere and Lennard-Jones particle systems  
by direct numerical simulation. 

\begin{figure}
\begin{center}
\includegraphics[width=6cm]{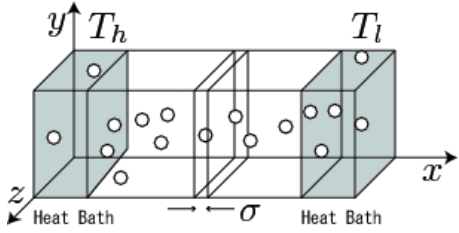}
\end{center}
\caption{Geometry of nonequilibrium particle simulation: The component of the kinetic part of thermal current is 
measured in the cross section of the system with the width $\sigma$ located at the center of the box.}
\label{fig:geom}
\end{figure}
In order to confirm the picture of the tail structure, 
we perform a nonequilibrium particle dynamics simulation. 
The geometry of the system is shown in Fig.~\ref{fig:geom}. Particles are confined in a rectangular 
parallelepiped box, whose size is denoted the $L_{x} \times L_{y} \times L_{z}$.
Heat bath regions are attached to both sides of the $x$-direction with a width $L_{b}$, where the kinetic 
temperature of particles is controlled by a Nos\'e-Hoover thermostat\cite{No84-1,No84-2,Ho85} with 
different temperatures $(T_{h},T_{l})$. The boundary conditions are as follows: In the  $x$-direction, we impose 
appropriate elastic wall potentials on both ends; in the $y$- and $z$-directions, periodic boundary conditions are 
imposed. The interaction potential for the inter-center distance $r$ of particles 
is taken to be a hard Hertzian potential\cite{Lo27}, $\phi(r) = Y \lvert \sigma -r \rvert^{5/2} \Theta(\sigma-r)$, 
where $Y$ is proportional to Young's modulus, which is taken to be $Y=10000 \epsilon \slash \sigma^{5/2}$ 
with an energy unit $\epsilon$, $\sigma$ is the characteristic length of the particles corresponding to a diameter of Hertzian particles, 
and $\Theta$ is the Heaviside step function. 
Empirically, this potential can reproduce the properties of a hard-sphere system. 
The dynamics of the particle is governed by Newtonian dynamics in a bulk system. 
Using heat baths with different temperatures, 
we realize a nonequilibrium steady thermal conducting state in the simulation. 
From previous studies, this setup enables a normal thermal conduction, which is characterized by 
the Fourier law (linear profile of local temperature) and a finite transport coefficient.\cite{SMYSI00,MSYI03,OYI05,OYI06}

\begin{figure}
\begin{center}
\includegraphics[width=8cm]{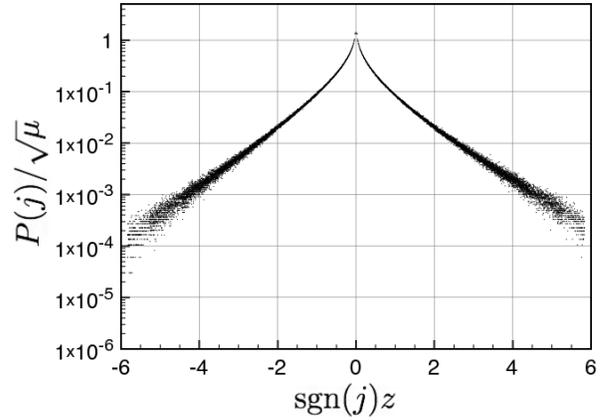}
\end{center}
\caption{Distribution of $\jc$, which is a parallel component of the thermal gradient:
The horizontal axis represents $\sgn(j) z$ and the vertical axis represents $ \ln ( P(j) \slash \sqrt{\mu}) $, where $z$ and $\mu $ are 
identical to definitions used in the equilibrium distribution eq.~(\ref{eq:eqdist}). Three different thermal gradient cases 
are superposed.}
\label{fig:xdist}
\end{figure}
In the following simulations, we take the following parameters: 
$L_{x} = L_{y} = 16 \sigma , L_{z} =  120 \sigma , \text{ and } L_{b} = 16 \sigma. $ 
The number of particles $N=3072.$ We take three different pairs of heat bath 
temperatures: $(T_{h},T_{l})=(10\epsilon, \epsilon), (20\epsilon, \epsilon), \text{ and } (40 \epsilon,\epsilon)$.
Discarding the initial relaxation steps, steady states are realized. Their local number density $n$ at the center is 
about $n \simeq 0.1 \slash \sigma^{3}$ in all cases, and their local kinematic temperature $T$ at the center 
are $T \simeq 6 \epsilon, 11\epsilon,$ and $23\epsilon$, respectively. 
The thermal gradients of actual steady states near the center are also calculated 
as $\partial T \slash \partial x \simeq 0.06\epsilon/\sigma, 0.12\epsilon/\sigma, 0.23\epsilon/\sigma$, respectively. 
Note that the log-thermal gradients of the actual states are $\partial \ln T \slash \partial x \simeq 0.01 \sigma^{-1}$ in all cases.

The distribution of $\jc$ is observed in the cross section of the system with a width $\sigma$ located at the center $[60\sigma, 61 \sigma)$. 
The results for about $3 \times 10^{6}$ samplings per case are shown in Fig.~\ref{fig:xdist}. 
We recognize that the distributions of $\jc$  with three different thermal gradients are well scaled with 
$\sgn(j) z$ and $P(j)\slash \sqrt{\mu}$ from the figure. 
This scaling collapse of the distribution for three different thermal gradients is consistent with the analysis based on the Boltzmann equation. 
In the Boltzmann analysis, the thermal gradient always appears in the form, $\partial \ln T \slash \partial x$, instead of $\partial T \slash \partial x$.\cite{CC70} In the present simulation, the log-thermal gradients are $0.01$ in all cases. 
Thus the scaling collapse means that the simulation data have the same dependency on the thermal gradient as the Boltzmann equation. 
The simulation also confirms that the distributions of the other two components are identical to the equilibrium 
distribution $P_{\mathrm{eq}}(j)$.

\begin{figure}
\begin{center}
\includegraphics[width=8cm]{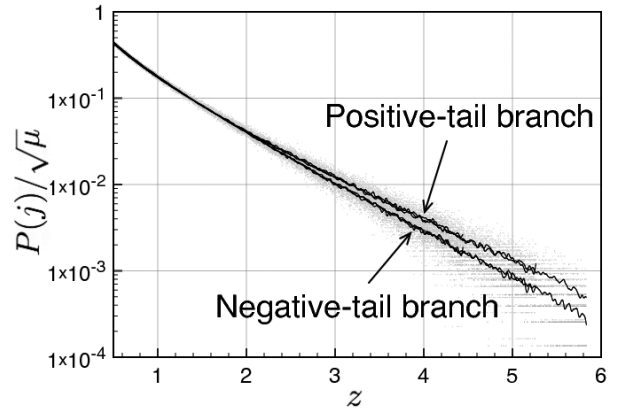}
\end{center}
\caption{Negative tails are flipped at $z=0$ and superposed over positive tails:
Gray scattered dots are original data and black lines are averaged over several bins for each data point of three different gradients.
Clearly, we observe two branches for negative and positive currents.}
\label{fig:tails}
\end{figure}
The tails of the distribution in Fig.~\ref{fig:xdist} are slightly skewed. 
To investigate the form of the tail in detail, we replot the data 
by flipping the negative tail at $z=0$ and superposing the positive tail
with statistical smoothing over several bins. The result is shown in Fig.~\ref{fig:tails}. 
Plots from three different thermal gradients clearly show the same curve again.
In addition, branches of positive and negative tails are clearly distinguished. 

\begin{figure}
\begin{center}
\includegraphics[width=8cm]{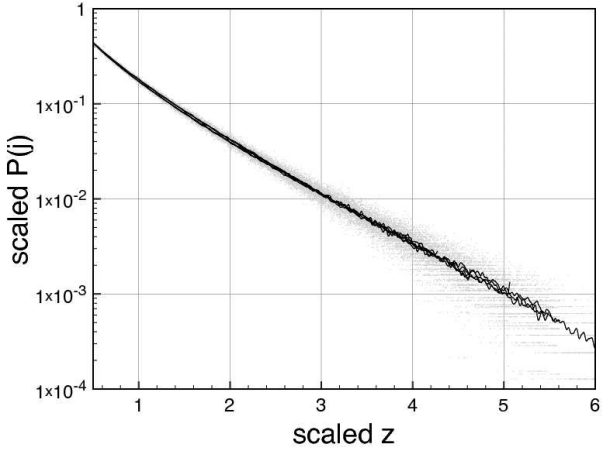}
\end{center}
\caption{Scaling plot of the negative and positive tails: 
Gray scattered dots are scaled original data and black lines are scaled branches for three different thermal gradients. Six curves 
are well collapsed into a single curve. 
}
\label{fig:scaling}
\end{figure}
If the conjecture that the tails of the distribution are identical to equilibrium distributions with different temperatures is true, 
we can expect another type of scaling from the form of the equilibrium distribution eq.~(\ref{eq:eqdist}):
\begin{equation}
z \to \dfrac{z}{\alpha_{\pm}} , \qquad \dfrac{P(j)}{\sqrt{\mu}} \to \alpha_{\pm}^{3/2}\dfrac{P(j)}{\sqrt{\mu}} 
\enspace ,
\end{equation}
where $\alpha_{\pm}$ is a rescaling factor of local temperature. For the positive tail, 
its temperature is given by $T_{\mathrm{sc}}^{>} = \alpha_{+} T$ and, for the negative tail, its temperature 
is given by $T_{\mathrm{sc}}^{<} = \alpha_{-} T $. These temperatures are called ``scaling'' temperatures.
The best fit of the above scaling is given by $\alpha_{\pm} = 1 \pm 0.04$ and the result is shown in Fig.~\ref{fig:scaling}.
This scaling factor is also related to the thermal gradient $\partial T \slash \partial x$ and the characteristic length $l$ as 
\begin{equation}
\alpha_{\pm} T \simeq T \pm \dfrac{\partial  T}{\partial x} l \enspace .
\end{equation}
$\alpha_{\pm} = 1 \pm 0.04$ and $\partial \ln T \slash \partial x = 0.01 \sigma^{-1}$ give the characteristic 
length $l$ as $l= 4 \sigma$. The mean free path $l_{\mathrm{f}}$ of the present situation is calculated 
as $l_{\mathrm{f}} = 1 \slash \sqrt{2} \pi n \sigma^{2} \simeq 2.25 \sigma$. Thus, the picture of 
nonequilibrium tails are consistent with the computational result.

Here, we use another definition of tail temperature.  
Calculation of the new defined temperature is easier than calculation of the scaling temperature: 
Local-equilibrium kinetic temperature $T$ is calculated by 
\begin{equation}
T = \dfrac{1}{3m} \langle \mathbf{p}^{2}\rangle_{\mathrm{ness}} \enspace ,
\end{equation}
where $\langle \dots \rangle_{\mathrm{ness}} $ means that the average is calculated in the local section in NESS. 
Decomposing this average into two parts with positive and negative $p_{x}$ ensembles, 
we can define the new temperature $T_{\mathrm{eff}}^{>} \,\,(T_{\mathrm{eff}}^{<})$ for the positive (negative) tail:
\begin{equation}
T_{\mathrm{eff}}^{>} = \dfrac{1}{3m} \langle \mathbf{p}^{2} \rangle_{\mathrm{ness}}^{p_{x} > 0} \text{ and }  T_{\mathrm{eff}}^{<} = \dfrac{1}{3m} \langle \mathbf{p}^{2} \rangle_{\mathrm{ness}}^{p_{x} < 0} \enspace .
\end{equation}
From the definition, the following relation exists: $T= (T_{\mathrm{eff}}^{>} + T_{\mathrm{eff}}^{<} )\slash 2$.
This temperature is a sort of ``effective'' temperature because of its difference from the scaling temperature. 
The difference comes from contributions near $j\sim 0$. 
For the present results, $T_{\mathrm{eff}}^{>} \slash T$ and $T_{\mathrm{eff}}^{<} \slash T$, 
which correspond to the alpha parameters of the scaling temperature, are $1\pm 0.0235(5), 1\pm 0.0284(4), \text{ and } 1\pm 0.0335(5), $ for $\partial T \slash \partial x \simeq 0.06 \epsilon \slash \sigma, \simeq 0.12 \epsilon \slash \sigma , \text{ and } 0.23 \epsilon \slash \sigma$, respectively. 
With increasing thermal gradient, the difference between the  two temperatures becomes much smaller.

For the Lennard-Jones particle system, a similar result is also observed. 
The Lennard-Jones interaction is taken to be as follows:
\begin{equation}
\phi(r) = 4 \varepsilon \left\{
  \left(\dfrac{r_{c}}{r}\right)^{12}-\left(\dfrac{r_{c}}{r}  \right)^{6}
  \right\}, 
\end{equation}
with a potential cutoff length $3 r_{c}$. 
Simulations with the same setups as the Hertzian particle system are conducted with the following parameters: 
$L_{x} =L_{y}=L_{z} = 15 r_{c}$, and $ L_{b}=2r_{c}$.  The number of particles is $N=843$. 
Distribution is measured at the center $[7r_{c},8r_{c})$. 
The average number density is $n\simeq 0.249 \slash r^{3}_{c}$.  Several temperature gradients, 
$(T_h,T_l)$ $=$ $(3.5\varepsilon,3.3\varepsilon)$ $(3.6\varepsilon,3.2\varepsilon)$ 
$(3.8\varepsilon,3.0\varepsilon)$ $(4.0\varepsilon,2.8\varepsilon)$ $(4.2\varepsilon,2.6\varepsilon)$ $(5.0\varepsilon,1.8\varepsilon)$, 
are examined. These parameters are chosen as the local kinetic temperature in 
the observation section becomes about $3.4\varepsilon$. 
From these parameters, the Lennard-Jones system is in a supercritical fluid phase. 
The simulation gives us a skewed nonequilibrium distribution of $\jc$. 
The temperatures of the tails also differ from the local equilibrium temperature. 
We can calculate the scaling and kinetic temperatures. 
Again, there are differences between both temperatures. 
\begin{figure}
\begin{center}
\includegraphics[width=7cm]{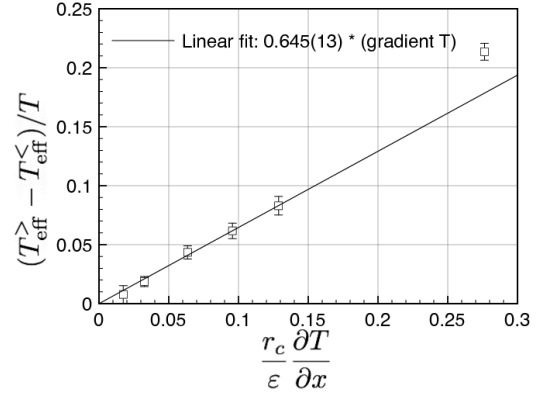}
\caption{Difference in effective kinetic temperature between both tails normalized by the local kinetic temperature 
plotted against the local temperature gradient for the Lennard-Jones particle system. }
\label{fig:LJtails}
\end{center}
\end{figure}
For the Lennard-Jones system, we plot the difference between $T_{\mathrm{eff}}^{>}$ and $T_{\mathrm{eff}}^{<}$ normalized by $T$ against several temperature gradients  in Fig.~\ref{fig:LJtails}. 
In the small-gradient region, a linear relation is observed. 
It seems that linearity is broken in a rather large thermal gradient. At this thermal gradient, 
the linearity of thermal transport is also broken.

Such kind of tail property, in which the negative and positive tails obey 
the equilibrium distributions with different temperatures 
asymptotically, are not special properties of particle systems. Recently, for the quantum thermal 
transport system of the harmonic chain, it is shown that the distribution function of heat 
has exponential tails with different temperatures.\cite{SD07}
For the stochastic lattice thermal-conduction model proposed by Giardin\`a \textit{et al.}\cite{GKR07}
an exact nonequilibrium distribution of microscopic thermal current through a bond connecting site $i$ and site $i+1$ can be calculated.
The result is 
\begin{align}
P_{\mathrm{ness}}^{\mathrm{GKR}}(j;T_{i},T_{i+1}) &= \dfrac{1}{\sqrt{2\pi T_{i} T_{i+1}}} \exp \left( - \frac{j}{4} \left( \frac{
1}{T_{i}} - \frac{1}{T_{i+1}} \right) \right) \nonumber \\
&\times \mathrm{K_{0}} \left( \dfrac{\lvert
 j \rvert}{4} \left(\dfrac{1}{T_{i}}+\dfrac{1}{T_{i+1}} \right)\right)
\enspace ,
\end{align}
where $T_{i}$ and $T_{i+1}$ are the local temperatures of subsequent sites. 
The equilibrium distribution of the Giardin\`a \textit{et al.}'s model with a temperature $T$ is expressed as
\begin{equation}
P_{\mathrm{eq}}^{\mathrm{GKR}}(j;T) = \dfrac{1}{\sqrt{2\pi} T } 
\mathrm{K_{0}} \left( \dfrac{\lvert
 j \rvert}{2} \dfrac{1}{T}\right)
\enspace .
\end{equation}
The nonequilibrium distribution has the same tail property, because of the 
asymptotic behavior of the modified Bessel function $\mathrm{K}_{0}(x) \sim e^{- x} \slash\sqrt{x}, \text{ as } x \to \infty$:
\begin{multline}
P_{\mathrm{ness}}^{\mathrm{GKR}}(j;T_{i},T_{i+1}) \\
\propto
\begin{cases}
\lvert j \rvert^{-1/2}\exp \left( - \dfrac{j}{2} \dfrac{1}{T_{i}} \right) & j \to \infty \\
\lvert j \rvert^{-1/2}\exp \left( - \dfrac{\lvert j \rvert}{2} \dfrac{1}{T_{i+1}} \right) & j \to -\infty
\end{cases}
\enspace .
\end{multline}
Therefore, the behavior of the tail, in which the negative and positive tails obey 
the equilibrium distributions with different temperatures 
asymptotically, is a universal property of 
the nonequilibrium steady thermal transporting system, at least, 
in the linear nonequilibrium regime, although the asymptotic form itself depends on 
the dimensionality of the momenta and details of the models.

In this study, we have investigated the microscopic distribution of the kinetic part of heat current by direct 
nonequilibrium numerical simulations of particle systems. The result shows that the positive and negative tails 
of the distribution for the parallel component to the thermal gradient have an asymptotic form of the equilibrium 
distribution with different temperatures. These temperatures differ from the local equilibrium temperature. 
The temperature of the positive tail, which corresponds to the normal current direction, is larger 
than that of the negative tail. In addition, we have found that these temperatures are identical to those of the places separated 
by a distance several times the mean free path scale. 
Two definitions of the temperature of the tail have been studied: One is the scaling temperature, 
which is accurate but difficult to obtain, and the other is the effective kinematic temperature, 
which is less accurate but easy to calculate. Both temperatures can capture of the tail structure well. 

The tail structure is not a special property of the particle system, but it may be a 
universal property of a normal thermal transport. 
For a nonlinear regime of thermal transport, the same tail structure of $\jc$ has 
already been observed in the Lennard-Jones particle system. Investigations of 
the potential part of thermal current that consists of  potential advection and work terms are now in progress. 

This work was partly supported by a Grant-in-Aid for Scientific Research (B) No. 19340110, 
a Grant-in-Aid for Young Scientists (B) No. 19740238
from the Ministry Education, Culture, Sports, Science and Technology Japan, and the Global Research Partnership of King Abdullah 
University of Science and Technology (KUK-I1-005-04).

\end{document}